\documentclass[twocolumn,preprintnumbers,showpacs,prb,amsmath,amssymb]{revtex4}

\usepackage{graphicx}
\usepackage{dcolumn}
\usepackage{bm}

\usepackage{hyperref}

\graphicspath{{Figures/}}

\begin{document}

\title{Single-shot readout of multiple nuclear spin qubits in diamond\\
 under ambient conditions}
\author{A. Dr\'eau$^{1}$}
\author{P. Spinicelli$^{1}$}
\author{J. R. Maze$^{2}$}
\author{J.-F.~Roch$^{3}$}
\author{V.~Jacques$^{1}$}
\email{vjacques@lpqm.ens-cachan.fr}
\affiliation{$^{1}$Laboratoire de Photonique Quantique et Mol\'eculaire, Ecole Normale Sup\'erieure de Cachan and CNRS UMR 8537, 94235 Cachan, France}
\affiliation{$^{2}$Facultad de F\'{i}sica, Pontificia Universidad Cat\'{o}lica de Chile, Santiago 7820436, Chile}
\affiliation{$^{3}$ Laboratoire Aim\'e Cotton, CNRS UPR 3321 and Universit\'e Paris-Sud, 91405 Orsay, France}

\begin{abstract}
We use the electronic spin of a single Nitrogen-Vacancy (NV) defect in diamond to observe the real-time evolution of neighboring single nuclear spins under ambient conditions. Using a diamond sample with a natural abundance of $^{13}$C isotopes, we first demonstrate high fidelity initialization and single-shot readout of an individual $^{13}$C nuclear spin. By including the intrinsic $^{14}$N nuclear spin of the NV defect in the quantum register, we then report the simultaneous observation of quantum jumps linked to both nuclear spin species, providing an efficient initialization of the two qubits. These results open up new avenues for diamond-based quantum information processing including active feedback in quantum error correction protocols and tests of quantum correlations with solid-state single spins at room temperature.
\end{abstract}


\maketitle

Nuclear spins are attractive candidates for solid-state quantum information storage and processing owing to their extremely long coherence time~\cite{Chuang_Science1997,Kane_Nature1998,Ladd_Nature2010}. However, since this appealing property results from a high level of isolation from the environment, it remains a challenging task to polarize, manipulate and readout with high fidelity individual nuclear spins~\cite{{Balestro_Nature2012}}. A promising approach to overcome this limitation consists in utilizing an ancillary single electronic spin to detect and control remote nuclear spins coupled by hyperfine interaction~\cite{McCamey_Science2010,Steger_Science2012,Dutt_Science2007,Neumann_Science2010,Robledo_Nature2011,Maurer_Science2012}. In this context, the NV defect in diamond has recently attracted considerable interest because its electronic spin can be polarized, coherently manipulated, and readout by optical means with long coherence times, even under ambient conditions~\cite{Balasubramanian_NatMater_2009}. The NV's electronic spin thus behaves as an ultrasensitive magnetometer at the nanoscale~\cite{Taylor2008}, providing a robust interface to detect and control nearby nuclear spins in the diamond lattice. This approach has been used in the past years to study the coherent dynamics of multi-spin systems~\cite{Childress_Science2006}, to perform universal quantum gates~\cite{Jelezko_PRL2004,Tono_Nature2012} and to develop few-qubits quantum registers, where single nuclear spins are used as quantum memories~\cite{Dutt_Science2007,Neumann_Science2008,Fuchs_NatPhys2011}. A second-long coherence time was recently demonstrated for a single $^{13}$C nuclear spin weakly coupled to a single NV defect in an isotopically purified diamond sample~\cite{Maurer_Science2012}. This result, combined with the ability to perform spin-photon entanglement~\cite{Togan_Nature2011} and two-photon interference from distant NV defects at low temperature~\cite{Sipahigil_PRL2012,Bernien_PRL2012}, makes single spins in diamond a promising building block for quantum repeaters and long-distance quantum communications.\\
\indent However, advanced quantum algorithms such as quantum error correction protocols require high fidelity initialization and single-shot readout over multiple qubits~\cite{Chuang}. Along the line of recent works directed towards this goal~\cite{Jiang_Science2009,Maurer_Science2012,Robledo_Nature2011,Neumann_Science2010}, we first report high fidelity single-shot readout of an individual $^{13}$C nuclear spin by using the electronic spin of a single NV defect as an ancillary qubit in a diamond sample with a natural abundance of $^{13}$C isotopes ($1.1\%$). Repetitive readout indicates a polarization lifetime exceeding seconds at moderate magnetic fields, which illustrates the robustness of the $^{13}$C nuclear spin state.  Then, we demonstrate efficient initialization of two nuclear spin qubits in a well-defined state by adding the intrinsic $^{14}$N nuclear spin of the NV defect in the quantum register.\\
\begin{figure}[h!]
\includegraphics[width = 8.8cm]{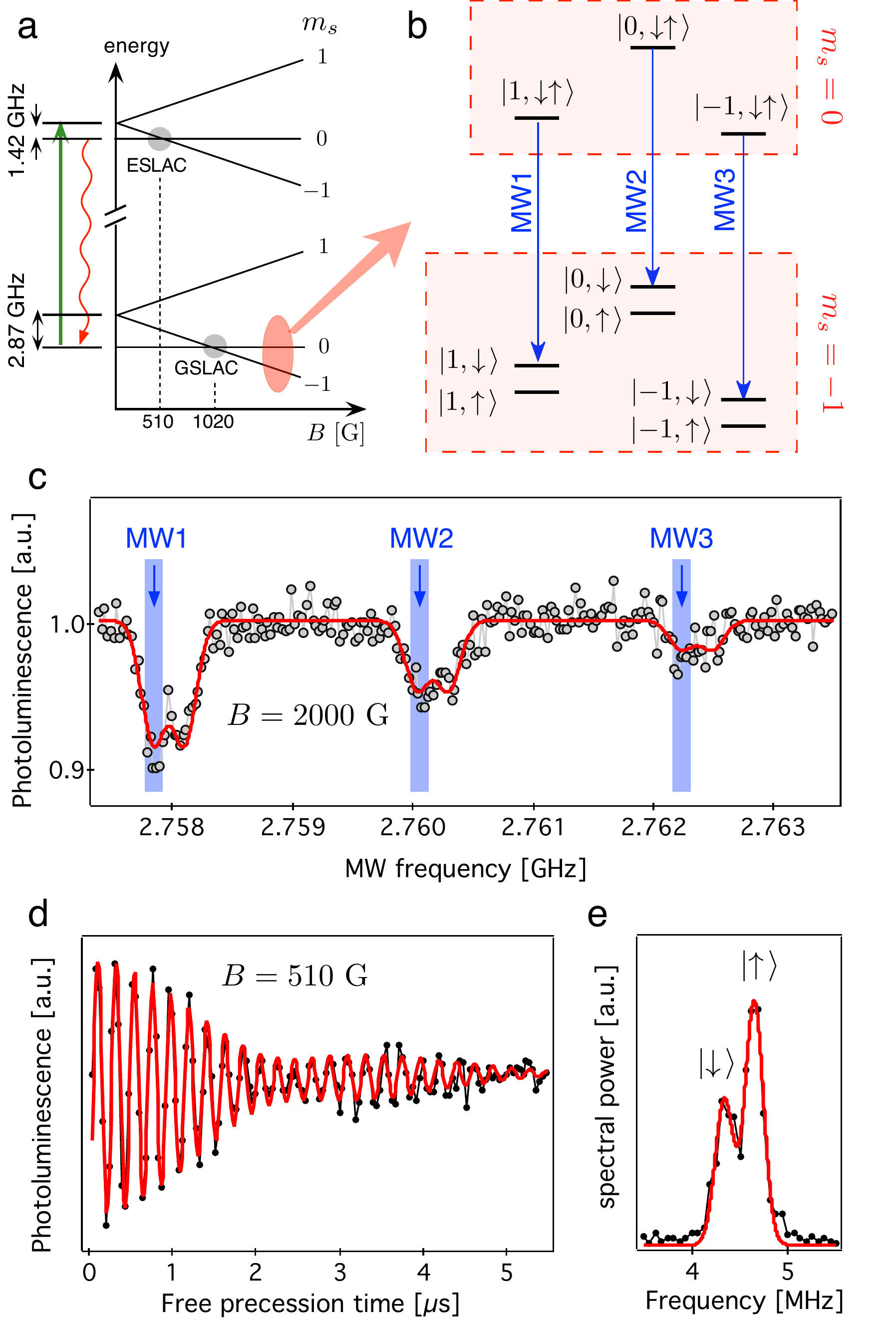}
\caption{Spin system. (a), Energy-level diagram of the NV defect as a function of the strength of a static magnetic field $B$ applied along the NV defect axis. Level anti-crossings in the ground state (GSLAC) and in the excited state (ESLAC) are highlighted. (b), Hyperfine structure of the $m_{s}=0$ and $m_{s}=-1$ electron spin manifolds for a NV defect coupled with its intrinsic $^{14}{\rm N}$ nuclear spin (nuclear spin projection $m_{I^{\rm (N)}}$) and with a nearby single $^{13}{\rm C}$ nuclear spin (nuclear spin projection $\uparrow$ or $\downarrow$). The hyperfine sublevels are denoted as $\left|m_{I^{{\rm (N)}}},\uparrow\downarrow\right.\rangle$ and the blue arrows indicate the microwave (MW) transitions used for single-shot readout measurements. (c), Optically detected ESR spectrum recorded for a magnetic field $B\approx 2000$~G. The $^{14}$N hyperfine interaction leads to a splitting of $2.16$ MHz between ESR frequencies associated with different $^{14}$N nuclear spin projections. These lines are further split by ${A}_{zz}=258\pm10$~kHz through hyperfine coupling with a nearby $^{13}$C. (d), FID signal of the NV defect electron spin recorded at the ESLAC ($B\approx 510$~G) showing a coherence time $T_{2}^{*}=2.9\pm0.1 \ \mu$s, in the range expected for a diamond sample with a natural abundance of $^{13}$C. (e)-Fourier transform of the FID signal showing a significant polarization ($\sim 40\%$) of the $^{13}{\rm C}$ nuclear spin in state $\left|\uparrow\right.\rangle$.}
\label{Fig1}
\end{figure}
The spin system considered in this study is depicted in Figs.~\ref{Fig1}(a) and (b). The electronic spin ($S=1$) of a single NV defect is coupled by hyperfine interaction with both its intrinsic $^{14}$N nuclear spin ($I=1$) and a neighboring $^{13}$C nuclear spin ($I=1/2$). A permanent magnet placed on a three-axis translation stage is used to apply a static magnetic field with controlled amplitude along the NV defect axis and the spin transition between the $m_{s}=0$ and $m_{s}=-1$ electron spin manifolds is coherently driven through microwave (MW) excitation. As shown in Fig.~\ref{Fig1}(c), the hyperfine structure of the spin system, recorded through pulsed-ESR spectroscopy~\cite{Dreau_PRB2011}, exhibits six nuclear-spin conserving transitions (see Supplementary Information). From this spectrum, recorded for a magnetic field magnitude $B=2000$~G, we extract the projected strength of the $^{13}$C hyperfine interaction $\mathcal{A}_{\|}=\mathcal{A}_{zz}=258\pm10$~kHz. We obtain further qualitative information of the hyperfine interaction through dynamic polarization measurements at the excited-state level anti-crossing (ESLAC), while applying a static magnetic field near $510$ G along the NV axis~\cite{Jacques_PRL2009}. The $^{13}$C polarization efficiency was estimated by using the Fourier transform of the free-induction decay (FID) signal measured by applying a Ramsey sequence $\frac{\pi}{2}-\tau-\frac{\pi}{2}$ to the NV defect electron spin. Figure~\ref{Fig1}(d)) shows the FID signal recorded at the ESLAC. Since the $^{14}$N nuclear spin is perfectly polarized, the characteristic beating is linked to the weakly coupled $^{13}$C nuclear spin. The Fourier transform of the FID signal indicates a relatively high polarization efficiency $\mathcal{P}=40\pm 10\%$, which suggests that the $^{13}$C quantization axis is close to the NV defect axis in both the ground and excited states~\cite{Gali_PRB2009}. The anisotropic component of the hyperfine tensor $\mathcal{A}_{ani}$ is therefore assumed to be much smaller than $\mathcal{A}_{zz}$. Since the $^{13}$C nuclear spin gets polarized in $\left|\uparrow\right.\rangle$, this measurement also provides unambiguous identification of each ESR frequency to a given nuclear spin state, $\left|\uparrow\right.\rangle$ or $\left|\downarrow\right.\rangle$~\cite{Smeltzer_NJP2011,Dreau_PRB2012}.

\begin{figure}[t]
\includegraphics[width=8.7cm]{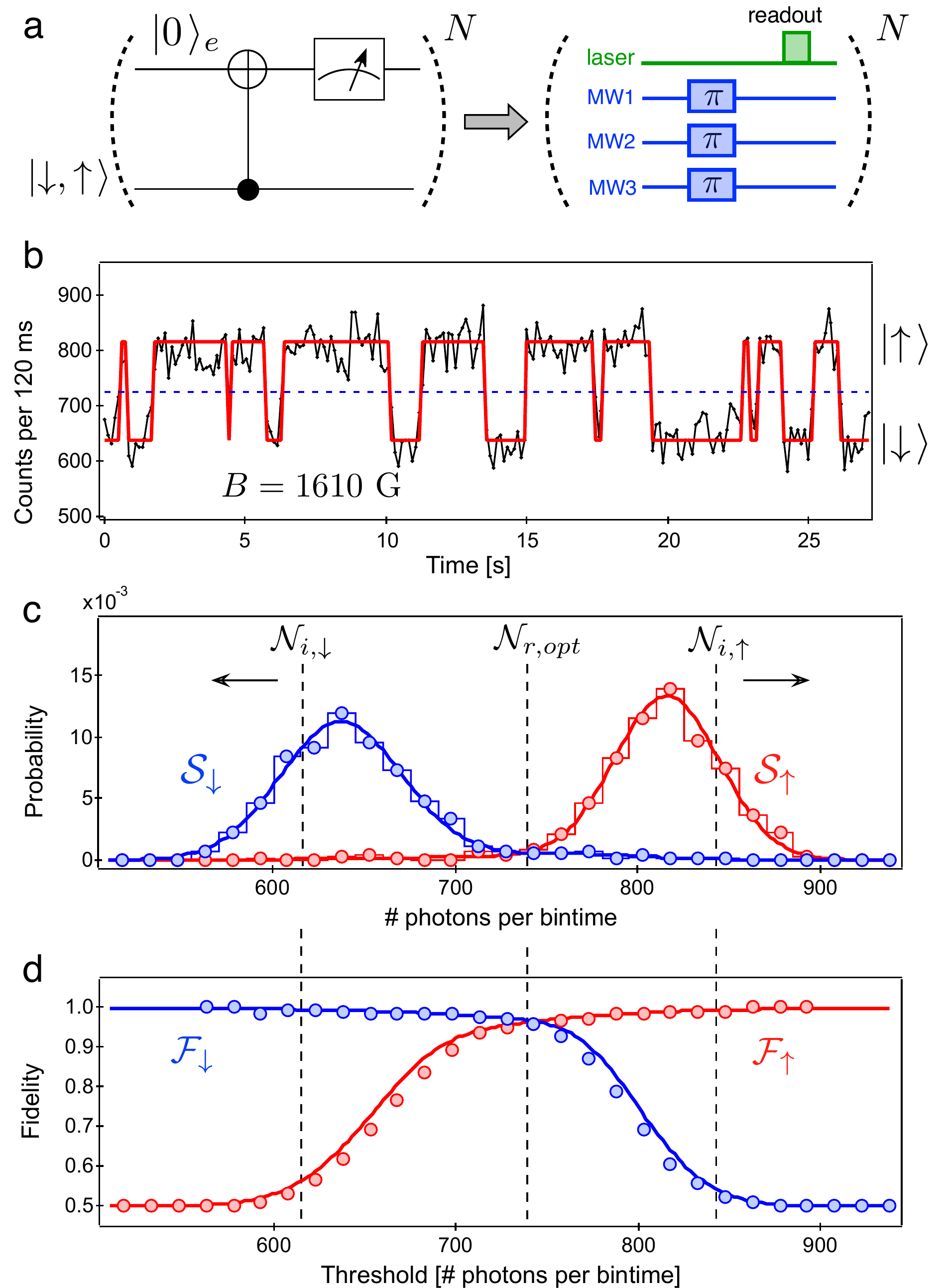}
\caption{Single-shot readout of a single $^{13}$C nuclear spin. (a), Logic diagram of the single-shot readout scheme and corresponding experimental sequence. For all experiments, the duration of the $\pi$-pulses is set to $4 \ \mu$s. The 300-ns laser pulse is used both for spin-state read-out and to achieve an efficient preparation of the NV defect electron spin in the $m_s=0$ sublevel ($\left|0\right.\rangle_{e}$) at each repetition of the sequence. (b), PL time trace showing quantum jumps of the $^{13}$C nuclear spin state. The solid line is a fit with a two states hidden Markov model from which the relaxation time $T_{1}$ of the nuclear spin state is extracted. (c), Normalized nuclear-spin dependent photon counting distributions $\mathcal{S}_{\uparrow(\downarrow)}$. The solid lines are data fitting with the formula given in the Supplementary Information. (d), Single-shot readout fidelity $\mathcal{F}_{\uparrow(\downarrow)}$ as a function of the readout threshold. The initialization thresholds $\mathcal{N}_{i,\downarrow}$,$\mathcal{N}_{i,\uparrow}$ and the optimized discrimination threshold $\mathcal{N}_{r,opt}$ are indicated with dashed lines. The solid lines are extracted from the fits in (c). We note that the initialization fidelity in state $\left|\downarrow\right.\rangle$ (resp. $\left|\uparrow\right.\rangle$) is given by $\mathcal{F}_{\downarrow}(\mathcal{N}_{i,\downarrow})$ (resp. $\mathcal{F}_{\uparrow}(\mathcal{N}_{i,\uparrow})$).}
\label{Fig2}
\end{figure}

\indent In the spirit of previous works~\cite{Neumann_Science2010,Maurer_Science2012}, projective single-shot detection of the $^{13}$C nuclear spin state is achieved by accumulating the NV defect photoluminescence (PL) while repeating the sequence depicted in Fig.~\ref{Fig2}(a). The NV defect electron spin is first initialized into the $m_s=0$ sublevel through optical pumping. A controlled not (CNOT) gate is then applied to induce an electron spin-flip conditioned on the $^{13}$C nuclear spin state. Finally, the resulting electronic spin state is optically readout by applying a $300$-ns laser pulse. This sequence is repeated many times in order to increase the signal to noise ratio. The CNOT gate is experimentally realized by applying narrowband MW $\pi$-pulses on the electronic spin, which selectively drive the ESR transition for a given $^{13}$C nuclear spin state, {\it e.g.} $\left|\downarrow\right.\rangle$. In order to take advantage of the full ESR contrast and to get rid off any quantum jumps linked to the $^{14}$N nuclear spin~\cite{Neumann_Science2010}, three MW sources are used for driving  simultaneously $^{13}$C nuclear spin state-selective transitions from each hyperfine sublevels linked to the $^{14}$N nucleus (Figs.~\ref{Fig1}(b) and (c)).

A typical PL time trace recorded while continuously repeating the sequence is shown in Fig.~\ref{Fig2}(b). For each data point, the PL signal is accumulated during $\tau_{b}=120$~ms, corresponding to approximatively 20000 repetitions of the readout sequence. The signal exhibits well-defined quantum jumps linked to the evolution of the $^{13}$C nuclear spin state. Indeed, when the nuclear spin is in state $\left|\downarrow\right.\rangle$, the CNOT gate flips the NV defect electron spin, $m_s=0\rightarrow m_s=-1$, and a low PL signal is observed ({\it dark} state) owing to spin-dependent PL of the NV defect. Conversely, when the nuclear spin is in state $\left|\uparrow\right.\rangle$, the electron spin remains in the $m_s=0$ sublevel at each repetition of the sequence and a high PL signal is observed ({\it bright} state). Nuclear spin flips are therefore evidenced in real-time as abrupt jumps between two distinct values of the PL signal. For a magnetic field $B=1610$~G applied along the NV defect axis, we infer the characteristic relaxation times of the $^{13}$C nuclear spin while applying repetitive readout $T_{1,\uparrow (bright)}=2.4\pm0.1$~s and $T_{1,\downarrow (dark)}=1.5\pm0.2$~s.

To estimate the readout fidelity, the $^{13}$C nuclear spin is first deterministically initialized in a given state through a single-shot readout measurement. By introducing an initialization threshold $\mathcal{N}_{i,\downarrow}$ (resp. $\mathcal{N}_{i,\uparrow}$), photon counting events such that $\mathcal{N}<\mathcal{N}_{i,\downarrow}$ (resp. $\mathcal{N}>\mathcal{N}_{i,\uparrow}$) are post-selected, corresponding to an initialization in state $\left|\downarrow\right.\rangle$ (resp. $\left|\uparrow\right.\rangle$). Using $\mathcal{N}_{i,\downarrow}=615$~counts and $\mathcal{N}_{i,\uparrow}=845$~counts, the initialization fidelity exceeds $99\%$ for both nuclear spin states (see Fig.~\ref{Fig2}(d)). We note that in principle the initialization fidelity can be chosen arbitrarily high by decreasing (resp. increasing) $\mathcal{N}_{i,\downarrow}$ (resp. $\mathcal{N}_{i,\uparrow}$), at the price of a high number of lost events. After successful initialization, a subsequent readout measurement is performed allowing to build the nuclear-spin dependent photon counting distributions $\mathcal{S}_{\uparrow(\downarrow)}$. As shown in Fig.~\ref{Fig2}(c), the distributions linked to each nuclear spin state can be clearly distinguished and the readout fidelities $\mathcal{F}_{\uparrow(\downarrow)}$ are defined as 
\begin{equation}
\mathcal{F}_{\downarrow}(\mathcal{N}_{r})=\frac{\int_{0}^{\mathcal{N}_{r}}\mathcal{S}_{\downarrow}(\mathcal{N})d\mathcal{N}}{\int_{0}^{\mathcal{N}_{r}}\mathcal{S}_{\downarrow}(\mathcal{N})d\mathcal{N}+\int_{0}^{\mathcal{N}_{r}}\mathcal{S}_{\uparrow}(\mathcal{N})d\mathcal{N}} 
\end{equation}
\begin{equation}
\mathcal{F}_{\uparrow}(\mathcal{N}_{r})=\frac{\int_{\mathcal{N}_{r}}^{\infty}\mathcal{S}_{\uparrow}(\mathcal{N})d\mathcal{N}}{\int_{\mathcal{N}_{r}}^{\infty}\mathcal{S}_{\downarrow}(\mathcal{N})d\mathcal{N}+\int_{\mathcal{N}_{r}}^{\infty}\mathcal{S}_{\uparrow}(\mathcal{N})d\mathcal{N}} \ ,
\end{equation}
where $\mathcal{N}_{r}$ is the readout threshold. For $\mathcal{N}_{r,opt}=735$~counts, corresponding to the maximum overlap between the two photon-counting distributions, we extract $\mathcal{F}_{\downarrow}=\mathcal{F}_{\uparrow}=96 \pm 1.2 \%$ (Fig.~\ref{Fig2}(d)). This fidelity could be significantly improved by increasing the collection efficiency with diamond photonic nanostructures~\cite{Babinec_NatNano2010}. In addition, the selective MW $\pi$-pulses used for the CNOT gate have a duration of $4 \ \mu$s, corresponding to a spectral width of $130$ kHz. Given the inhomogeneous linewidth of the ESR signal combined with the hyperfine coupling strength, the $\pi$-pulses are therefore not perfectly selective leading to a decreased contrast of the projective measurement, which degrades the readout fidelity. This limitation could be overcome by using a CVD-grown diamond sample isotopically enriched with $^{12}\textrm{C}$ atoms, in which the inhomogeneous dephasing rate of the NV defect electron spin can reach few kHz~\cite{Maurer_Science2012,Balasubramanian_NatMater_2009}.

\begin{figure}[b]
\includegraphics[width=8.7cm]{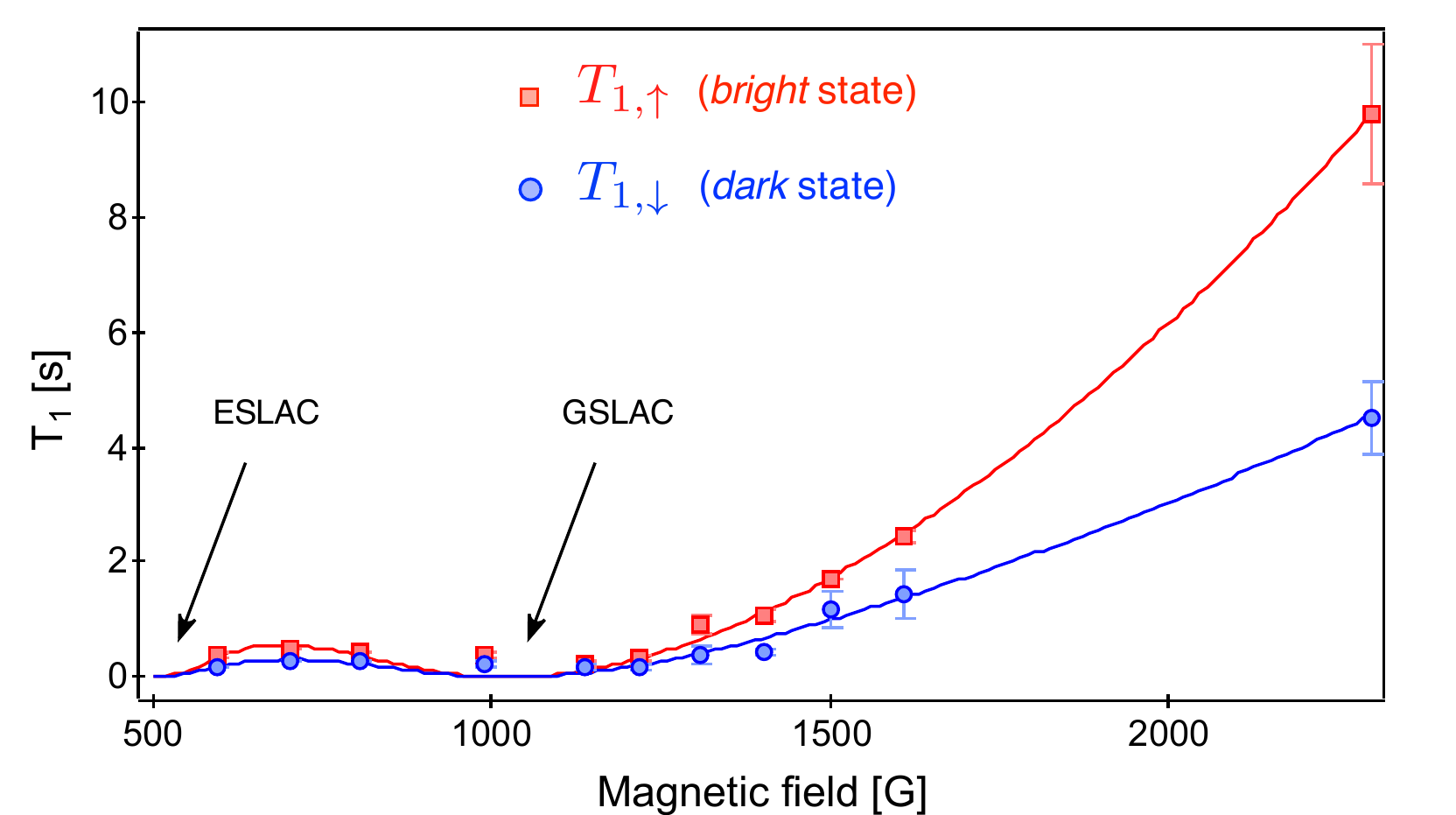}
\caption{Nuclear spin relaxation time $T_{1}$ as a function of the strength of a magnetic field applied along the NV defect axis for the {\it bright} state (red) and the {\it dark} state (blue). The solid lines are data fitting with a simple model including nuclear spin flips induced by the transverse component of the hyperfine tensor and electron-nuclear spin flip-flops at the GSLAC and ESLAC (see Supplementary Information).}
\label{Fig3}
\end{figure}

\begin{figure*}[t]
\includegraphics[width=16.5cm]{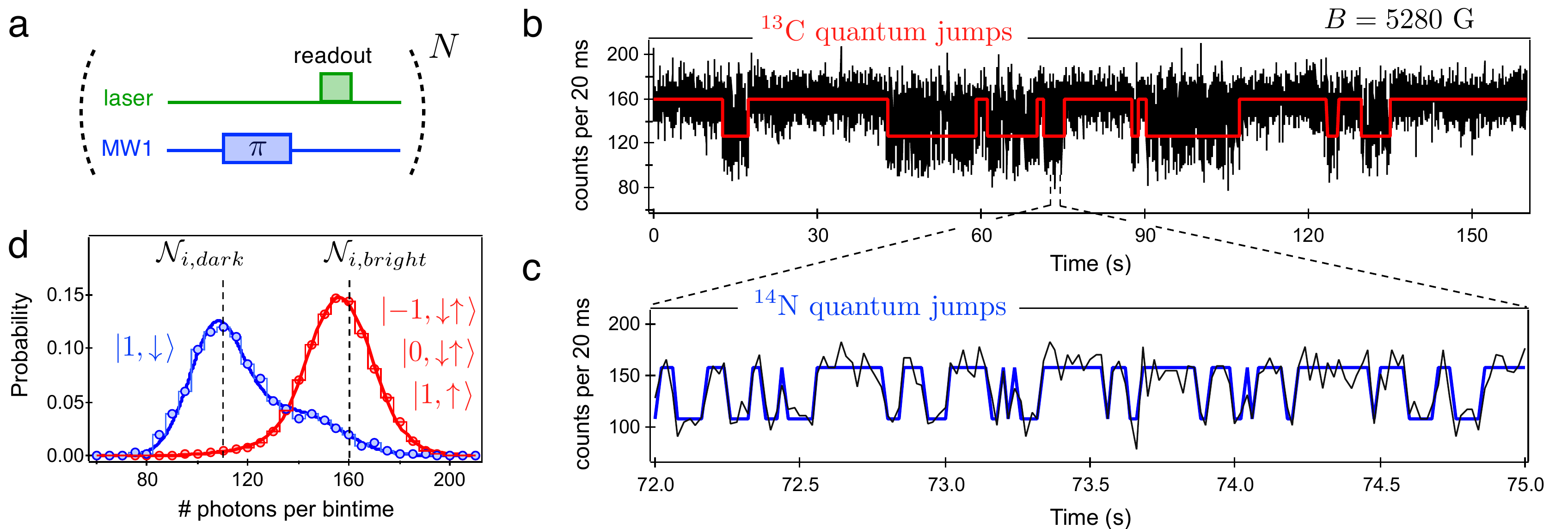}
\caption{Initialization and single-shot readout of a two nuclear spin qubit quantum register. (a), Experimental sequence. (b),(c), PL time trace recorded by continuously repeating the sequence with a magnetic field $B=5280$~G applied along the NV defect axis. Each data point corresponds to approximatively 3000 repetitions of the readout sequence (20 ms). Quantum jumps linked to (b) the weakly coupled $^{13}$C nuclear spin and to (c) the intrinsic $^{14}$N nuclear spin of the defect can be distinguished. (d), Normalized nuclear-spin dependent photon counting distributions measured with the initialization thresholds set to $\mathcal{N}_{i,dark}=110$~counts and $\mathcal{N}_{i,bright}=160$~counts. The projective readout fidelity is optimized for a discrimination threshold $\mathcal{N}_{r,opt}=135$~counts.}
\label{Fig4}
\end{figure*}

We now study the processes leading to nuclear spin depolarization. For a magnetic field $B$ applied along the NV defect axis ($z$), the ground-state spin Hamiltonian reads $\mathcal{H}=\mathcal{H}_{0}+\gamma_{n}B\hat{I}_{z}+\mathbf{\hat{S}}\cdot{\mathcal A}\cdot\mathbf{\hat{I}}$, where $\mathcal{H}_{0}$ includes both the pure electronic spin terms and the hyperfine interaction with the intrinsic $^{14}$N nuclear spin, $\gamma_{n}$ is the $^{13}$C gyromagnetic ratio and $\mathcal{A}$ its hyperfine tensor. In the secular approximation, this Hamiltonian simplifies to
\begin{equation}
\mathcal{H}=\mathcal{H}_{0}+\gamma_{n}B\hat{I}_{z}+\mathcal{A}_{zz}\hat{S}_{z}\hat{I}_{z}+\frac{\mathcal{A}_{ani}}{2}\left[ e^{-i\phi}\hat{S}_{z}\hat{I}_{+}+e^{+i\phi}\hat{S}_{z}\hat{I}_{-}\right] 
\label{Hamilto1}
\end{equation}
where $\mathcal{A}_{ani}=(\mathcal{A}_{zx}^{2}+\mathcal{A}_{zy}^{2})^{1/2}$, $\tan\phi = \mathcal{A}_{zy}/\mathcal{A}_{zx}$ and $\hat{I}_{\pm}=\hat{I}_{x}\pm i\hat{I}_{y}$. The anisotropic component $\mathcal{A}_{ani}$ of the hyperfine tensor therefore induces nuclear spin flips, leading to depolarization at a rate 
\begin{equation}
\gamma_{1}=\frac{1}{T_1}\propto  \frac{\mathcal{A}_{ani}^2}{\mathcal{A}_{ani}^2+(\mathcal{A}_{zz}-\gamma_{n}B)^2} \ .
\label{T1}
\end{equation}
Considering this process at the main source of depolarization, the nuclear spin relaxation time might exhibit a quadratic dependence with the applied magnetic field. The experimental results depicted in Figure~\ref{Fig3} confirm this behavior at high fields, while two drops can be observed around $B\sim 510$~G and $B\sim 1020$~G, corresponding to level anti-crossings in the excited state and in the ground state, respectively~\cite{Fuchs_PRL2011} (Fig.~\ref{Fig1}(a)). Around such magnetic field strengths, the secular approximation is not valid and additional electron-nuclear spin flip-flop terms $\mathcal{A}_{\perp}[\hat{S}_{-}\hat{I}_{+}+\hat{S}_{+}\hat{I}_{-}]/2$ need to be added to the Hamiltonian~\cite{Jacques_PRL2009,Gali_PRB2009}, where $\mathcal{A}_{\perp}=(\mathcal{A}_{xx}+\mathcal{A}_{yy})/2$. As shown in Fig.~\ref{Fig3}, the experimental data are well fitted by a simple model including depolarization induced by the anisotropic hyperfine interaction and spin mixing at the level anti-crossings (see Supplementary Information for details). We note that the {\it bright} state always exhibits a longer relaxation time than the {\it dark} state. Furthermore, this effect is independent on the nuclear spin state ($\left|\uparrow\right.\rangle$ or $\left|\downarrow\right.\rangle$) used as control state in the CNOT gate. When the {\it dark} state is detected, a shorter nuclear spin lifetime is always observed because in this case the system spends on average more time in the $m_s=-1$ electronic spin sublevel, for which the anisotropic component of the hyperfine tensor induces nuclear spin flips. 

According to equation~(\ref{T1}), a long nuclear spin lifetime can be observed either for a $^{13}$C nuclear spin with a weak anisotropic component of the hyperfine interaction, {\it i.e.} placed on a lattice site with a small angle with respect to the NV defect axis, or for an applied magnetic field such that $\gamma_{n}B\gg (\mathcal{A}_{zz},\mathcal{A}_{ani})$. In a diamond sample with a natural abundance of $^{13}\textrm{C}$ isotope ($1.1\%$), the ESR linewidth is on the order of $200$~kHz, which puts a limit to the weakest detectable hyperfine coupling strength in conventional ESR spectroscopy (Fig. 1(c)). Apart from the $^{13}$C nuclear spin studied in detail in this work, quantum jumps were also observed for a $^{13}$C coupling strength $\mathcal{A}_{zz}=380\pm10$~kHz (lattice site O in Ref.~[28]) with a much shorter relaxation time (see Supplementary Information for details). For stronger hyperfine coupling strengths, no quantum jumps could be observed for magnetic fields up to $5000$~G. The probability to find weakly coupled $^{13}$C nuclear spins would be significantly improved by using isotopically purified diamond samples~\cite{Maurer_Science2012}. However, we note that the speed of the single-shot readout measurement decreases with the $^{13}$C coupling strength owing to the required spectral selectivity of the quantum logic.

Finally, we demonstrate single-shot readout in a two-qubit register by including the intrinsic $^{14}$N nuclear spin of the NV defect. For this experiment, the CNOT gate is performed with a single narrowband MW $\pi$-pulse which selectively drive the ESR transition for a given state of the register, {\it e.g.} state $\left|1,\downarrow\right.\rangle$ (Fig. 1(b) and Fig. 4(a)). The PL time trace then exhibits quantum jumps linked to both nuclear spin species, which can be easily distinguished because their characteristic relaxation times differ by orders of magnitude (Figs. 4(b) and (c)). Indeed, although the $^{14}$N nuclear spin shares its symmetry axis with the NV defect ($\mathcal{A}_{ani}=0$), its relaxation time is only a few tens of milliseconds because a strong hyperfine contact interaction in the NV defect excited-state $\mathcal{A}_{\perp}\approx 40$~MHz induces fast electron-nuclear spin flip-flops~\cite{Neumann_Science2010}. From the nuclear-spin dependent photon counting distributions, we infer that the two nuclear spin qubits can be initialized into state $\left|1,\downarrow\right.\rangle$ ({\it dark} state) with a fidelity higher than $98\%$ by using an initialization threshold $\mathcal{N}_{i,dark}=110$~counts (Figs. 4(d)). We note that any state of the register could be deterministically prepared and readout by changing the frequency of the MW used for the CNOT gate. From the overlap between the photon counting distributions, we extract a projective readout fidelity $\mathcal{F}=83\pm2\%$, limited by the $^{14}$N nuclear spin relaxation time. This value could be significantly improved by increasing the magnetic field strength in order to decouple more efficiently the $^{14}$N nuclear spin from the electron spin dynamics~\cite{Neumann_Science2010}.

The reported initialization and single-shot readout of two nuclear spin qubits combined with well-established techniques of coherent manipulation within the quantum register~\cite{Jelezko_PRL2004,Tono_Nature2012} pave the way towards tests of quantum correlations with solid-state single spins at room temperature~\cite{Pfaff_Arxiv} and implementations of simple quantum error correction protocols~\cite{Chuang}.\\

\noindent {\it Aknowledgements.} The authors acknowledge P.~Bertet, J. Wrachtrup and M. Lecrivain for fruitful discussions and experimental assistance. This work was supported by the Agence Nationale de la Recherche (ANR) through the projects D{\sc iamag}, A{\sc dvice} and Q{\sc invc}. J.R.M. acknowledges support from Conicyt Fondecyt, Grant No.11100265, and US Air Force Grant FA9550-12-1-0214.


\begin{widetext}
\vspace{0.5cm}
\section*{SUPPLEMENTARY INFORMATION}
\subsection{Experimental methods}
\subsubsection{Experimental setup} 
We study native NV defects hosted in a commercial [100]-oriented high-purity diamond crystal grown by chemical vapor deposition (Element6) with a natural abundance of ${^{13}}$C isotopes ($1.1\%$). Individual NV defects are optically isolated at room temperature using a confocal microscope. A laser operating at $532$ nm wavelength is focused onto the diamond sample through a high numerical aperture oil-immersion microscope objective (Olympus, $\times 60$, NA=1.35) mounted on xyz-piezoelectric scanner (MCL, Nano-PDQ375). The red-shifted NV defect PL is collected by the same objective and spectrally filtered from the remaining excitation laser with a dichroic filter and a bandpass filter (Semrock, 697/75 BP). The collected PL is then directed through a 50-$\mu$m-diameter pinhole and focused onto a silicon avalanche photodiode (Perkin-Elmer, SPCM-AQR-14) operating in the single-photon counting regime. Laser pulses are produced with an acousto-optical modulator (MT200-A0.5-VIS) with a characteristic rising time of $10$~ns. For all experiments, the optical pumping power is set at $300 \ \mu$W, corresponding to the saturation power of the NV defect radiative transition.
 
\subsubsection{ESR spectroscopy}
The NV defect ground state has an electronic spin $S=1$ that can be efficiently polarized into its $m_s=0$ sublevel through optical pumping~\cite{MansonPRB2006}. In addition, the PL intensity is significantly higher ($\sim 30 \%$) when the $m_{s}=0$ state is populated allowing the detection of electron spin resonances (ESR) on a single NV defect by optical means~\cite{GruberScience1997}.

Coherent manipulation of the NV defect electron spin is performed by applying a microwave field through a copper microwire directly spanned on the diamond surface. Electron spin resonance (ESR) spectroscopy is performed through repetitive excitation of the NV defect with a resonant microwave $\pi$-pulse followed by a $300$-ns read-out laser pulse~\cite{DreauPRB2011}. ESR spectra are recorded by continuously repeating this sequence while sweeping the $\pi$-pulse frequency and recording the PL intensity. The microwave power is adjusted in order to set the $\pi$-pulse duration to $4 \ \mu$s, as verified by recording electron spin Rabi oscillations. In this conditions, the ESR linewidth is given by the inhomogeneous dephasing rate of the NV defect electron spin which is on the order of $200$~kHz for a diamond sample with a natural abundance of ${^{13}}$C isotope ~\cite{MizuochiPRB2009}. 

\begin{figure}[b] 
\begin{center}
\includegraphics[width=7.5cm]{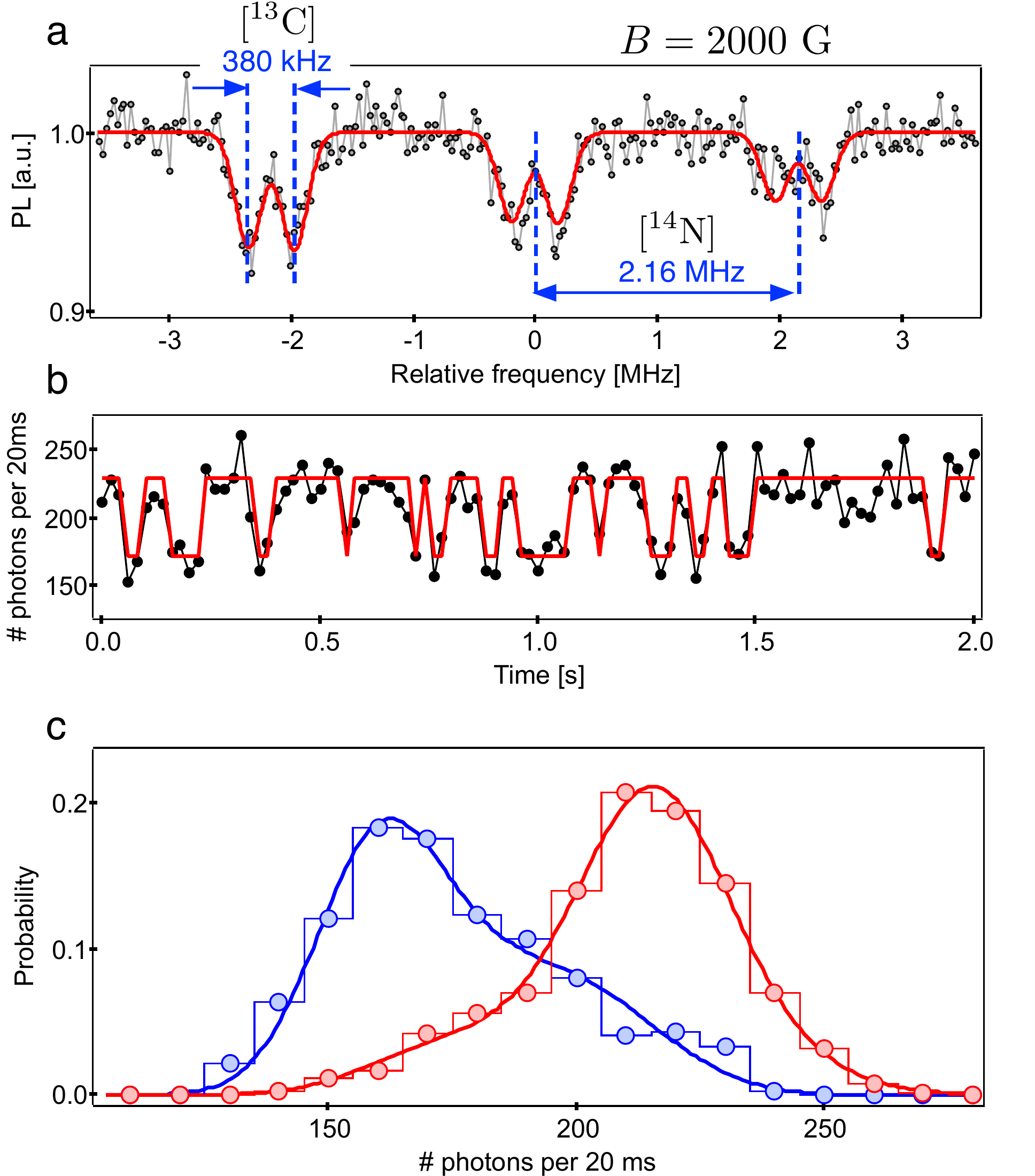}
\caption{Single-shot readout of a $^{13}$C nuclear spin with a hyperfine coupling strength of $380$~kHz. (a), Optically detected ESR spectrum recorded for a single NV defect coupled to a $^{13}$C nuclear spin with a coupling strength $\mathcal{A}_{zz}=380\pm10$~kHz. The solid line is a fit with Gaussian functions. (b), PL time trace recorded by continuously repeating the readout sequence shown in Fig. 2(a) of the main text, with a magnetic field $B=2000$~G applied along the NV defect axis. Each data point corresponds to approximatively 3000 repetitions of the readout sequence (20 ms). The solid line is a fit with a two states hidden Markov model. (c), Normalized nuclear-spin dependent photon counting distributions obtained  with the initialization thresholds set to $\mathcal{N}_{i,\downarrow}=150$~counts and $\mathcal{N}_{i,\uparrow}=240$~counts. With this values the initialization fidelity is $94\%$. }
\label{FigS1}
\end{center}
\end{figure}

The ESR spectrum of a single NV defect coupled with a nearby ${^{13}}$C nuclear spin shows six nuclear-spin conserving transitions (see Fig. 1(c) of the main text and Fig.~\ref{FigS1}(a)). Indeed, hyperfine interaction with the intrinsic $^{14}$N nuclear spin ($I=1$) leads to a splitting of $\mathcal{A}_{\rm N}=2.16$ MHz between ESR frequencies associated with different $^{14}$N nuclear spin projections~\cite{SmeltzerNJP2011,DreauPRB2012}. These lines are further split through hyperfine interaction with the ${^{13}}$C nuclear spin (Fig.~\ref{FigS1}(a)). Even at high magnetic fields, the $^{14}$N nuclear spin populations are unbalanced owing to dynamic nuclear spin polarization induced by optical pumping~\cite{JacquesPRL2009,SmeltzerPRA2009} (see Fig. 1(c) of the main text and Fig.~\ref{FigS1}(a)). We note that this effect is responsible for the short polarization time of the $^{14}$N nuclear spin under optical illumination~\cite{NeumannScience2010}.

\subsubsection{Magnetic field alignment}
A permanent magnet mounted on a xyz-translation stage is used to apply a static magnetic field along the NV defect axis. Preliminary alignment of the field is done by optimizing the PL intensity because any off-axis components of the magnetic field quench the NV defect PL~\cite{Epstein2005}. The field alignment is then more precisely realized by measuring the sum $\Sigma$ of the resonance frequencies $\nu_{+1}$ and $\nu_{-1}$, linked to the transitions $m_s=0\rightarrow m_s=+1$ and $m_s=0\rightarrow m_s=-1$, respectively. For a perfectly aligned magnetic field, $\Sigma=\nu_{+1}+\nu_{-1}=2D$, where $D$ is the zero-field splitting. In our experiments, this criteria is completed with a precision of about 100 kHz, corresponding to a magnetic field alignment with a precision better than $0.2^{\circ}$ for a magnetic field of $B=2000$~G.

\subsection{Single-shot readout of a $^{13}$C nuclear spin with a hyperfine coupling strength of $380$~kHz}

As indicated in the main text of the manuscript, quantum jumps were also observed for a $^{13}$C coupling strength $\mathcal{A}_{zz}=380\pm10$~kHz, as shown in Figure~\ref{FigS1}(b). For a magnetic field $B=2000$~G applied along the NV axis, the characteristic relaxation times of the $^{13}$C nuclear spin are $T_{1,\uparrow ({\it bright})}=69\pm4$~ms and $T_{1,\downarrow ({\it dark})}=35\pm2$~ms. From the overlap between the nuclear-spin dependent photon counting distributions (Fig.~\ref{FigS1}(c)), we infer a projective readout fidelity $\mathcal{F}=77\pm3 \%$. 

\subsection{Nuclear-spin dependent photon counting distributions}

In this section we describe how the histograms in Figure 2(c) of the main paper are obtained and how the statistics of the counted photons in our single-shot readout process is modeled. In the following the $^{13}$C nuclear spin states are denoted as {\it bright} and {\it dark}, corresponding to a high and a low photon-counting signal, respectively. These states correspond to either $\left|\uparrow\right.\rangle$ or $\left|\downarrow\right.\rangle$ depending on the nuclear-spin conserving ESR transition used for the CNOT gate. \\
\begin{figure}[h]
\begin{center}
\includegraphics[width=11cm]{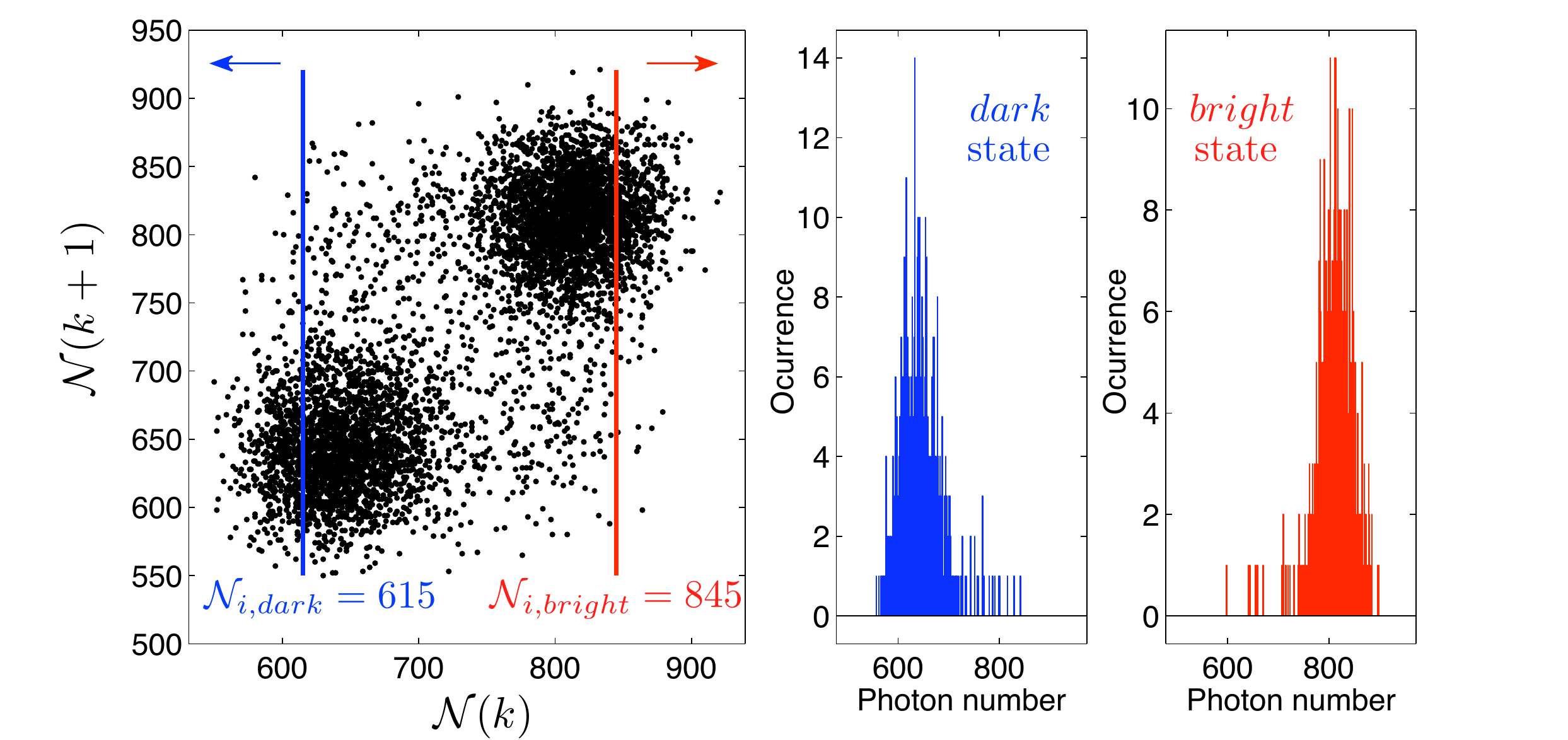}
\caption{ (Left) Distribution of consecutive single-shot measurements. (Right) Histograms of photon counting distributions for {\it dark} and {\it bright} states by using $\mathcal{N}_{i,{\it dark}} = 615$~counts and $\mathcal{N}_{{\it bright}} = 845$~counts.}
\label{fig:SC1}
\end{center}
\end{figure}

In order to obtain the nuclear-spin dependent photon counting histograms, we first consider the distribution of two consecutive measurements, $\mathcal{N}(k)$ and $\mathcal{N}(k+1)$, from the PL time trace obtained by applying continuously the single-shot readout sequence. As it can be seen on Figure~\ref{fig:SC1}, this distribution is highly concentrated in two regions representing both nuclear spin states, {\it dark} and {\it bright}. We define the threshold value $\mathcal{N}_{i,{\it dark}}$ (resp. $\mathcal{N}_{i,{\it bright}}$) on the measurement $\mathcal{N}(k)$ to declare that the nuclear spin state is initialized in the {\it dark} (resp. {\it bright}) state. Then, for all measurements that satisfy $\mathcal{N}(k)\leq \mathcal{N}_{i,{\it dark}}$ (resp. $\mathcal{N}(k)\geq \mathcal{N}_{i,{\it bright}}$), a histogram is constructed from the set of points $\{ \mathcal{N}(k+1) \}$ (see right panel in Figure~\ref{fig:SC1}). The lower (resp. larger) the $\mathcal{N}_{i,{\it dark}}$ (resp. $\mathcal{N}_{i,{\it bright}}$) threshold, the smaller the initialization error but the larger the uncertainty as we obtain fewer points to built the histograms. As a good compromise between these two effects, we choose $\mathcal{N}_{i,{\it dark}} = 615$~counts and $\mathcal{N}_{{\it bright}} = 845$~counts. Finally, each distribution is binned in intervals of $15$ counts to construct the nuclear-spin dependent photon counting histograms shown in Fig. 2(c) in the main paper.

We now explain how these histograms are modeled. As described in section D of this supplementary information, the flipping rate of the nuclear spin depends mainly on the anisotropic part of the hyperfine interaction between the electronic spin and the $^{13}$C nuclear spin. This causes an increase of the flip-flop rate when the electronic spin of the NV center is in state $m_s=\pm1$. As a consequence, the {\it dark} state always presents a smaller relaxation time $T_1$ than the {\it bright} state when the readout sequence is applied continuously.

Therefore we model the flip-flop events with a two-rate poissonian distribution. The time $Y_D$ (resp. $Y_{B}$) the nuclear spin spends on the {\it dark} (resp. {\it bright}) state distributes exponential with rate $\lambda_D$ (resp. $\lambda_B$). Following a similar procedure to that described on ref.~\cite{MaurerScience2012}, we model the number of photons detected over a measurement time $T$ with a random variable $Z$, by considering the statistics of the photons associated with each nuclear spin state when there is no flip, one flip and two flips over the measurement time.

When there is no-flip, the number of photons is given by the random variable,
\begin{eqnarray}
Z_I =  X_I \ , \ I=\{B,D\}
\end{eqnarray}
where $X_{I}$ is a random variable that distributes normal, $N_{I}(X_{I}=x_i)$, with mean $\mu_{I}=\mu_D$ (resp. $\mu_{I}=\mu_B$) and variance $\sigma_{I}^2=\sigma_D^2$ (resp. $\sigma_{I}^2=\sigma_B^2$) if the nuclear spin state is {\it dark} (resp. {\it bright}). 
The distribution of the number of photons when there is no-flip $f^0_{Z_I}(z)$ can then be calculated by taking the derivative with respect to $z$ of the cumulative distribution,
\begin{eqnarray}
f_{Z_I}^0(z) = \frac{\partial}{\partial z} \int_{x_i<z} N_I (x_i)P(Y_I>T) dx_i = N_I(z)e^{-\lambda_IT}.
\end{eqnarray}

When there is one flip during the measurement time $T$, we model the number of photons by the random variable,
\begin{eqnarray}
Z_{I} = \frac{Y_I}{T}X_I + \frac{T-Y_I}{T}X_{I'},
\end{eqnarray}

where $Y_I$ distributes exponentially with constant $\lambda_I$. The set of indexes $\{I, I'\}$ denotes the state of the nuclear spin and can be either $\{D,B\}$ or $\{B,D\}$. The distribution is given by
\begin{eqnarray}
f_{Z_I}^1(z) = \frac{\partial}{\partial z} \int \limits_{\frac{t}{T}x_I + \left(1-\frac{t}{T} \right)x_{I'}<z} N_I(x_I)N_{I'}(x_{I'})P(Y_I=t)P(Y_{I'}>T-t) dx_Idx_{I'}dt
\end{eqnarray}
\begin{eqnarray}
f_{Z_I}^1(z) & = & \int_0^1du  \frac{\lambda_ITe^{-\lambda_ITu} e^{-\lambda_{I'}T (1-u)}}{\sqrt{2\pi[ u^2\sigma_I^2 + (1-u)^2\sigma_{I'}^2]}}\exp\left\{-\frac{[z-(u\mu_I + (1-u)\mu_{I'})]^2}{2[u^2\sigma_I^2 + (1-u)^2\sigma_{I'}^2]} \right\}.\nonumber \\
\end{eqnarray}
Note that when $\lambda_I=\lambda_{I'}=\lambda$, the distribution becomes $f_Z^1(z) = P_I^1\int_0^1 duN(\mu,\sigma^2)$, where $P^1_I = \lambda T e^{-\lambda T}$, $N(\mu, \sigma^2)$ is the Normal distribution with mean $\mu$ and variance $\sigma^2$, $\mu = u\mu_I + (1-u)\mu_{I'}$ and $\sigma^2 = \sigma_I^2 u^2 + (1-u)^2\sigma_{I'}^2$.

Similarly, we model the two-flip case by a random variable
\begin{eqnarray}
{Z_I} &=& \frac{Y_I}{T}X_I + \frac{Y_{I'}-Y_I}{T}X_{I'} + \frac{T-Y_{I'}}{T}X_{I}\\
&=& \left(1-\frac{Y_{I'}-Y_I}{T} \right)X_I + \frac{Y_{I'}-Y_I}{T}X_{I'}
\end{eqnarray}
with distribution
\begin{eqnarray}
f_{Z_I}^2(z) & = & \int_0^1du  \frac{\lambda_IT\lambda_{I'}T(1-u)e^{-\lambda_IT} e^{+u(\lambda_IT-\lambda_{I'}T)}} {\sqrt{2\pi[ u^2\sigma_I^2 + (1-u)^2\sigma_{I'}^2]}}\exp\left\{-\frac{[z-(u\mu_I + (1-u)\mu_{I'})]^2}{2[u^2\sigma_I^2  + (1-u)^2\sigma_{I'}^2]} \right\}\nonumber \\
\end{eqnarray}
Note that when $\lambda_I=\lambda_{I'}=\lambda$, the distribution $f_Z^2(z) = P_I^2\int_0^1 du2(1-u)N(\mu,\sigma^2)$, where $P^2_I = (\lambda T)^2/2 e^{-\lambda T}$, $\mu = u\mu_I + (1-u)\mu_{I'}$ and $\sigma^2 = \sigma_I^2 u^2 + (1-u)^2\sigma_{I'}^2$.

Finally, we fit the experimental photon counting distributions shown in Fig 2(c) of the main paper to
\begin{eqnarray}
\mathcal{S}_{I}(z) = f_{Z_I}^0(z) + f_{Z_I}^1(z) + f_{Z_I}^2(z) \ ,
\end{eqnarray} 
where the fitting parameters are the average photon counting numbers $\mu_D$ and $\mu_B$, the flip-flop probabilities $T\lambda_D$ and $T\lambda_B$ for the nuclear spin in state {\it dark} and {\it bright}, respectively. On the other hand, it is known that three level systems with a metastable state exhibit super-poissonian character~\cite{Kim:PRA1987,Molski:CPL2009}. As the photons associated with the dark distribution involve the passage of the electron throught the metastable singlet state ${^1A_1}$, the variance of the dark distribution $\sigma_D^2$ is left as a fitting parameter, meanwhile the variance of the bright distribution is set to $\sigma_B^2 = \mu_B$. As a result we obtain $\mu_D = 637$~counts,  $\mu_D = 816$~counts,  $T\lambda_D = 0.084$ and $T\lambda_B = 0.0485$, leading to $T_{1,dark} =1.43$~s and $T_{1,bright} = 2.47$~s in fair agreement with our experimental results ($T_{1,dark} =1.5\pm0.2$ s and $T_{1,bright} =2.4\pm0.1$ s). 

\subsection{Nuclear-spin depolarization processes}

In this section, we discuss the depolarization processes of the $^{13}$C nuclear spin during the
multiple repetition of the single-shot readout sequence.

\subsubsection{System Hamiltonian}

The system consists of a single NV defect coupled by hyperfine interaction with its intrinsic $^{14}$N nuclear spin and a nearby $^{13}$C nuclear spin. Since we focus on the evolution of the $^{13}$C nucleus, we do not consider the interaction terms linked to the $^{14}$N.  Assuming a magnetic field $B$ perfectly aligned along the NV defect axis, denoted as the $z$-axis, the system Hamiltonian, in both the ground and excited states, reads  
\begin{equation}
\mathcal{H}^{(i)}= \mathcal{H}_{e}^{(i)} +\mathcal{H}_{n}^{(i)} +\mathcal{H}_{e-n}^{(i)} 
\end{equation}
with the index $i$ refering either to the ground state $(gs)$ or to the excited state $(es)$, and 
\begin{equation}
\left\{
\begin{array}{rl}
	\mathcal{H}_e^{(i)} &= D^{(i)} \hat{S}_z^2 +\gamma_e B \hat{S}_z \\
	\mathcal{H}_n^{(i)} & = \gamma_n B \hat{I}_z\\
	\mathcal{H}_{e-n}^{(i)} & = \hat{\mathbf{S}} \cdot \mathcal{A}^{(i)} \cdot \hat{\mathbf{I}} \ ,
\end{array}
\right.
\end{equation}
where $D^{(i)}$ is the zero-field splitting of the NV defect electronic spin - $D^{(es)} \simeq 1.42$~GHz~\cite{Fuchs_PRL_2008} and $D^{(gs)} \simeq 2.87$~GHz - , $\gamma_e \simeq 2.80$ MHz.G$^{-1}$ and $\gamma_n \simeq 1.07$ kHz.G$^{-1}$ are respectively the gyromagnetic ratio of the electronic spin and of the $^{13}$C nuclear spin, and $\mathcal{A}^{(i)}$ is the hyperfine tensor. 

Given the small value of the hyperfine coupling strength considered in the main text of the manuscript ($\mathcal{A}_{\parallel} \simeq 258$ kHz), the $^{13}$C must be located few lattice sites away from the NV defect~\cite{SmeltzerNJP2011,DreauPRB2012}. However, no correspondence with a specific lattice site of the diamond matrix is available owing to the current accuracy of {\it ab initio} calculations of the electronic spin wave function~\cite{SmeltzerNJP2011}. In the following, we neglect the contact term of the hyperfine interaction and thus assume a purely point-like dipolar interaction. The hyperfine tensor is therefore considered identical in the ground and in the excited states $\mathcal{A}^{(i)} = \mathcal{A}$. Furthermore, the significant $^{13}$C polarization at the excited state level anti-crossing suggests the polar angle between the $^{13}$C lattice site and the NV defect axis is small~\cite{GaliPRB2009}. This is further supported by the observed long polarization time of the $^{13}$C nuclear spin while applying repetitive readout. Thus, we can neglect the terms of the hyperfine interaction proportional to $S_{\pm} I_{\pm}$~\cite{Cohen} and the Hamiltonian can be approximated as 
\begin{eqnarray}
\mathcal{H}^{(i)} &\simeq& \mathcal{H}_e^{(i)} + \mathcal{H}_n^{(i)} + \mathcal{A}_{zz} \hat{S}_z \hat{I}_z +  \frac{\mathcal{A}_{ani}}{2} [\hat{S}_+ \hat{I}_z e^{-i \phi} + \hat{S}_- \hat{I}_z e^{+i \phi} ] \\
&+&\frac{\mathcal{A}_{\perp}}{2} [\hat{S}_+ \hat{I}_-  + \hat{S}_- \hat{I}_+ ] + \frac{\mathcal{A}_{ani}}{2} [\hat{S}_z \hat{I}_+ e^{-i \phi} + \hat{S}_z \hat{I}_- e^{+i \phi} ]
\label{hamiltonian}
\end{eqnarray}
where $\mathcal{A}_{ani}=\sqrt{\mathcal{A}_{zx}^2+\mathcal{A}_{zy}^2}$, $\tan \phi = \mathcal{A}_{zy}/\mathcal{A}_{zx}$ and $\mathcal{A}_{\perp} = (\mathcal{A}_{xx}+\mathcal{A}_{yy})/2=-\mathcal{A}_{zz}/2$ because the hyperfine tensor is traceless for a pure dipolar interaction. The last two terms of this Hamiltonian are responsible for the undesired $^{13}$C nuclear spin flips. 

\subsubsection{Nuclear spin flip processes}

The first nuclear spin-flip term is proportional to the perpendicular component of the hyperfine tensor $\mathcal{A}_{\perp}$, and connects the state $\mid -1 \rangle_e \mid \uparrow \rangle$ [resp. $\mid 0 \rangle_e \mid \uparrow \rangle$] with the state $\mid 0\rangle_e \mid \downarrow \rangle$ [resp. $\mid +1 \rangle_e \mid \downarrow \rangle$], as depicted on Figure \ref{Fig_depolar}(a). The nuclear spin-flip rate induced by this coupling is proportional to the transition probability between the two interacting states. Since the detuning between $\mid 0 \rangle_e \mid \uparrow \rangle$ and $\mid +1 \rangle_e \mid \downarrow \rangle$ is always larger than the one between $\mid -1 \rangle_e \mid \uparrow \rangle$  and $\mid 0\rangle_e \mid \downarrow \rangle$, the main depolarization rate linked to the perpendicular component of the hyperfine tensor can be expressed as 
\begin{equation}
	\gamma_{\perp}^{(i)} = \frac{1}{T_{1 \perp}^{(i)}} \propto  \frac{2 \mathcal{A}_{\perp}^2}{2 \mathcal{A}_{\perp}^2+(D^{(i)}-\gamma_e B)^2} \ .
\end{equation}
As shown on Figure \ref{Fig_depolar}(c), this rate becomes significant when the system gets close to level anti-crossings, {\it i.e.} for $D^{(i)}\approx \gamma_e B$, occurring around $B\sim 1020$~G and $B\sim 510$~G, in the ground state and in the excited state, respectively.

\begin{figure}[t]
\includegraphics[width= 8cm]{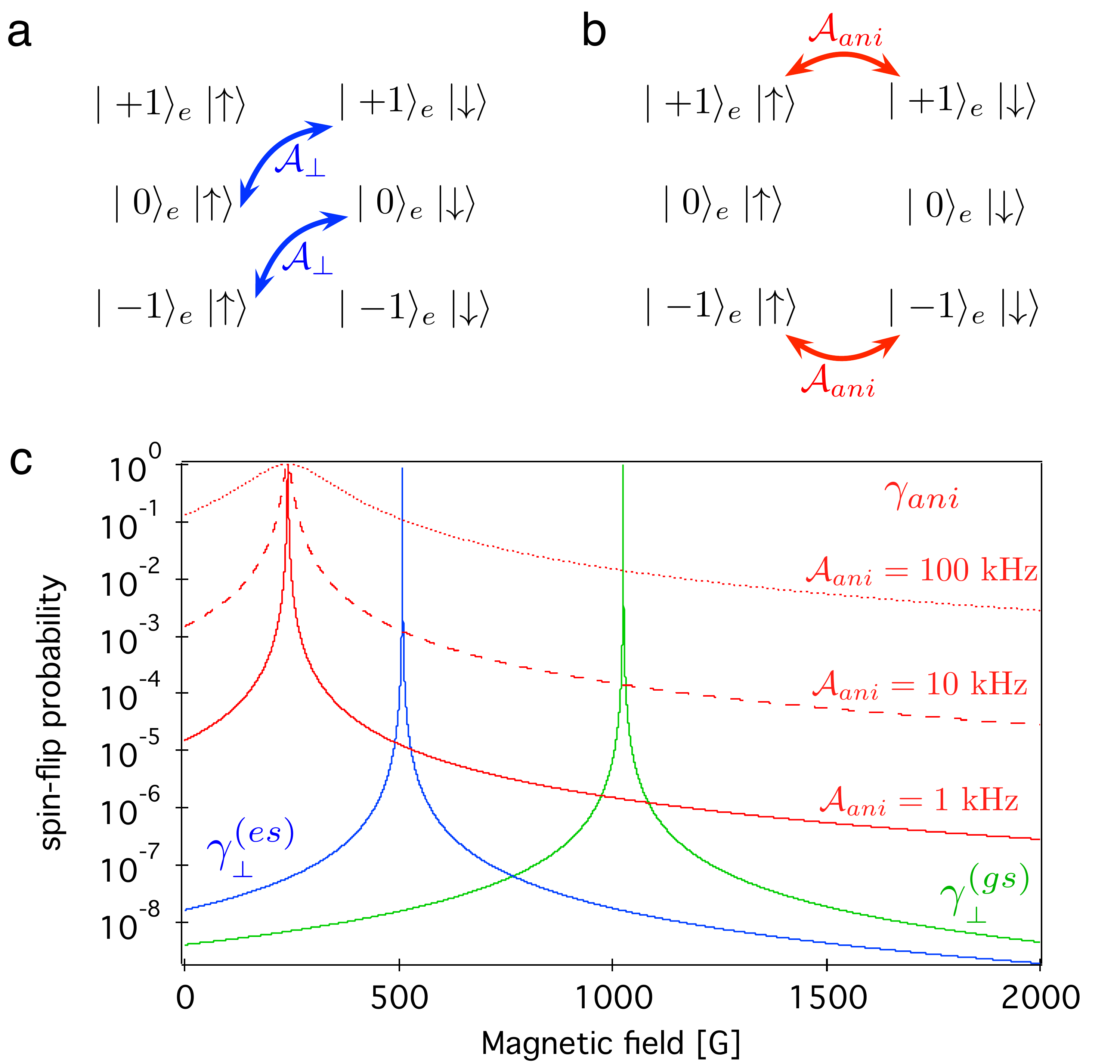}
\caption{Schemes illustrating the coupling between the different states $\mid m_s \rangle_e \mid \uparrow \downarrow \rangle$, resulting either from the perpendicular component $\mathcal{A}_{\perp}$ (a) or from the anisotropic component (b) of the hyperfine interaction. (c), Nuclear spin-flip rates linked to the anisotropic hyperfine coupling (red curves), and to the perpendicular hyperfine coupling in the ground (green curve) and excited state (blue curve). The parameters are set to $\mathcal{A}_{zz} = 258$ kHz, $\gamma_n = 1.07$ kHz.G$^{-1}$, $\gamma_e = 2.80$ MHz.G$^{-1}$, $D^{(gs)} = 2.87$ GHz and $D^{(es)} = 1.42$ GHz and $\mathcal{A}_{ani} = 1$ kHz (red solid line)$, 10$ kHz (red dashed line) and $100$ kHz (red dotted line).}
\label{Fig_depolar}
\end{figure}

The second nuclear spin-flip term in equation (\ref{hamiltonian}) is proportional to the anisotropic component of the hyperfine tensor $\mathcal{A}_{ani}$ (Fig.~\ref{Fig_depolar}(b)). For all the experiments reported in the main text, we use nuclear-spin conserving transitions between the $m_{s}=0$ and $m_{s}=-1$ electron spin manifolds. Since the anisotropic hyperfine interaction couples the states $\mid -1\rangle_e \mid \uparrow \rangle$ and $\mid -1\rangle_e \mid \downarrow \rangle$, the nuclear-spin-flip rate is therefore given by 
\begin{equation}
	 \gamma_{ani} = \frac{1}{T_{1,ani}} \propto \frac{\mathcal{A}_{ani}^2}{\mathcal{A}_{ani}^2+(\mathcal{A}_{zz}-\gamma_n B)^2} \ .
\end{equation}
The evolution of this rate with the magnetic field amplitude $B$ is shown on Figure \ref{Fig_depolar}(c) for different values of $\mathcal{A}_{ani}$. Even for a value as small as $\mathcal{A}_{ani}=1$~kHz, the anisotropic component of the hyperfine interaction is always the dominant depolarization process, except near the two level anti-crossings. This is supported by the experimental results depicted in Figure 3 of the main text, which indicate that the {\it {\it bright}} state, corresponding to the electronic spin in state $\mid 0\rangle_e$, always exhibits a longer relaxation time than the {\it {\it dark}} state. 

\subsubsection{$T_1$ evolution with the magnetic field}

The evolution of the nuclear spin relaxation time versus the magnetic field is depicted in Figure 3 of the main text. At a given magnetic field amplitude, $T_{1,\uparrow ({\it bright})}$ (red dots) and $T_{1,\downarrow ({\it dark})}$ (blue squared dots) are inferred from a fit to a PL time trace showing quantum jumps with a two-state Hidden Markov Model~\cite{NeumannScience2010,NR}.

We simply consider the total depolarization rate $\gamma_1$ as a weighted average of the three depolarization rates $\gamma_{ani}$, $\gamma_{\perp}^{(gs)}$ and $\gamma_{\perp}^{(es)}$, leading to the formula 
\begin{equation}
	\gamma_1=\frac{1}{T_1} \simeq \alpha_{ani} \gamma_{ani}+\alpha_{\perp}^{(gs)} \gamma_{\perp}^{(gs)} + \alpha_{\perp}^{(es)} \gamma_{\perp}^{(es)} \ ,
\label{fitting_formula}
\end{equation} 
where \{$\alpha_{ani},\alpha_{\perp}^{(gs)},\alpha_{\perp}^{(es)}$\} are coefficients linked to the optical pumping power, the intrinsic photophysical parameters of the NV defect and the time the system spends in each state $\mid i\rangle_e \mid \uparrow\downarrow \rangle$ during the single-shot readout sequence. The solid lines in Figure 3 represent a fit of the experimental results using equation (\ref{fitting_formula}) with \{$\alpha_{ani},\alpha_{\perp}^{(gs)},\alpha_{\perp}^{(es)}$\} as fitting parameters and setting $\mathcal{A}_{ani}=10$~kHz. Although this highly simplified model does not allow to extract quantitative information, it reproduces fairly the general trend of the experimental data. The development of a more precise model would require to introduce the NV defect dynamics under optical pumping, including ionization of the defect in the neutral charge state NV$^{0}$ (Ref.~\cite{MaurerScience2012}).

We note that the spin relaxation time could be significantly enhanced by aligning the magnetic field along the $^{13}$C hyperfine field rather than the NV axis, leading to $\mathcal{A}_{ani}\approx 0$. This could be realized for a $^{13}$C placed at a lattice site with a small polar angle  with respect to the NV defect axis in order to avoid any significant electronic spin mixing which degrade the ESR contrast.

\end{widetext}


\begin{thebibliography}{99}

\bibitem{Chuang_Science1997}
N. A. Gershenfeld and I. L. Chuang, Science {\bf 275}, 350-356 (1997).

\bibitem{Kane_Nature1998}
B. E. Kane, Nature {\bf 393}, 133-137 (1998).

\bibitem{Ladd_Nature2010}
T. D. Ladd, F. Jelezko, R. Laflamme, Y. Nakamura, C. Monroe, and J. L. OÕBrien, Nature {\bf 464}, 45-53 (2010).

\bibitem{Balestro_Nature2012}
R. Vincent, S. Klyatskaya, M. Ruben, W. Wernsdorfer, and F. Balestro, Nature {\bf 488}, 357-360 (2012) 

\bibitem{McCamey_Science2010}
D. R. McCamey, J. Van Tol, G. W. Morley, and C. Boehme, Science {\bf 330}, 1652-1656 (2010).

\bibitem{Steger_Science2012}
M. Steger {\it et al.}, Science {\bf 336}, 1280-1283 (2012).

\bibitem{Dutt_Science2007}	 
M. V. G. Gurudev Dutt {\it et al.}, Science {\bf 316}, 1312-1316 (2007). 

\bibitem{Robledo_Nature2011}
L. Robledo, L. Childress, H. Bernien, B. Hensen, P. F. A. Alkemade, and R. Hanson, Nature {\bf 477}, 574-578 (2011).

\bibitem{Neumann_Science2010}
P. Neumann {\it et al.}, Science {\bf 329}, 542-544 (2010). 

\bibitem{Maurer_Science2012}
P. C. Maurer {\it et al.},  Science {\bf 336}, 1283-1286 (2012).

\bibitem{Balasubramanian_NatMater_2009}
G. Balasubramanian {\it et al.}, Nature Mater. \textbf{8}, 383-387 (2009).

\bibitem{Taylor2008} 
J. M. Taylor {\it et al.}, Nature Phys. \textbf{4}, 810-816 (2008).

\bibitem{Childress_Science2006}
L. Childress {\it et al.}, Science {\bf 314}, 281-284 (2006).

\bibitem{Jelezko_PRL2004}
F. Jelezko, T. Gaebel, I. Popa, M. Domhan, A. Gruber, and J. Wrachtrup, Phys. Rev. Lett. {\bf 93}, 130501 (2004).

\bibitem{Tono_Nature2012}
T. van der Sar {\it et al.}, Nature {\bf 484}, 82-86 (2012).

\bibitem{Neumann_Science2008}
P. Neumann {\it et al.}, Science {\bf 320}, 1326-1329 (2008).

\bibitem{Fuchs_NatPhys2011}
G. D. Fuchs, G. Burkard, P. V. Klimov, and D. D. Awschalom, Nature Phys. {\bf 7}, 789-793 (2011).

\bibitem{Togan_Nature2011}
E. Togan {\it et al.}, Nature {\bf 466}, 730-734 (2010).

\bibitem{Sipahigil_PRL2012}
A. Sipahigil {\it et al.}, Phys. Rev. Lett. {\bf 108}, 143601 (2012).

\bibitem{Bernien_PRL2012}
H. Bernien, L. Childress, L. Robledo, M. Markham, D. Twitchen, and R. Hanson, Phys. Rev. Lett. {\bf 108}, 043604 (2012).

\bibitem{Chuang}
M. A. Nielsen and I. L. Chuang, {\it Quantum Computation and Quantum Information} (Cambridge Univ. Press, 2000).

\bibitem{Jiang_Science2009}
L. Jiang {\it et al.}, Science {\bf 326}, 267-272 (2009).
	
\bibitem{Dreau_PRB2011}
A. Dr\'eau {\it et al.}, Phys. Rev. B {\bf 84}, 195204 (2011).

\bibitem{Jacques_PRL2009}
V. Jacques {\it et al.}, Phys. Rev. Lett. {\bf 102}, 057403 (2009).

\bibitem{Gali_PRB2009}
A. Gali, Phys. Rev. B {\bf 80}, 241204 (2009).

\bibitem{Smeltzer_NJP2011}
B. Smeltzer, L. Childress, and A. Gali, New J. Phys. {\bf 13}, 025021 (2011).

\bibitem{Dreau_PRB2012}
A. Dr\'eau, J. R. Maze, M. Lesik, J.-F. Roch, and V. Jacques, Phys. Rev. B {\bf 85}, 134107 (2012).

\bibitem{Babinec_NatNano2010}
T. Babinec, M. Khan, Y. Zhang, J. R.  Maze, P. R. Hemmer, and M. Loncar, Nature Nano. {\bf 5}, 195-199 (2010).

\bibitem{Fuchs_PRL2011}
G. D. Fuchs {\it et al.}, Phys. Rev. Lett. {\bf 101}, 117601 (2008).

\bibitem{Pfaff_Arxiv}
W. Pfaff {\it et al.}, Nature Phys. doi:10.1038/nphys2444 (2012).

\end{thebibliography}

\begin{thebibliography}{99}

\bibitem{MansonPRB2006}
N. B. Manson, J. P. Harrison, and M. J. Sellars, Nitrogen-vacancy center in diamond: Model of the electronic structure and associated dynamics. {\it  Phys. Rev. B} {\bf 74,} 104303 (2006). 

\bibitem{GruberScience1997} 
A. Gr$\ddot{\rm u}$ber, A. Drabenstedt, C. Tietz, L.  Fleury, J. Wrachtrup, and C. von Borczyskowski, Scanning confocal optical microscopy and magnetic resonance on single defect centers. \textit{Science} \textbf{276,} 2012-2014 (1997).


\bibitem{DreauPRB2011}
A. Dr\'eau {\it et al.}, Avoiding power broadening in optically detected magnetic resonance of single NV defects for enhanced dc magnetic field sensitivity. {\it Phys. Rev. B} {\bf 84}, 195204 (2011).

\bibitem{MizuochiPRB2009}
N. Mizuochi \textit{et al.}, Coherence of single spins coupled to a nuclear spin bath of varying density. {\it Phys. Rev. B} \textbf{80}, 041201(R) (2009).

\bibitem{SmeltzerNJP2011}
B. Smeltzer, L. Childress, and A. Gali, $^{13}$C hyperfine interactions in the nitrogen-vacancy centre in diamond. {\it New J. Phys.} {\bf 13}, 025021 (2011).

\bibitem{DreauPRB2012}
A. Dr\'eau, J. R.  Maze, M. Lesik, J.-F. Roch, and V. Jacques, High-resolution spectroscopy of single NV defects coupled with nearby $^{13}$C nuclear spins in diamond. {\it Phys. Rev. B} {\bf 85}, 134107 (2012).

\bibitem{JacquesPRL2009}
V. Jacques \textit{et al.}, Dynamic Polarization of Single Nuclear Spins by Optical Pumping of Nitrogen-Vacancy Color Centers in Diamond at Room Temperature. \textit{Phys. Rev. Lett.} \textbf{102}, 057403 (2009).

\bibitem{SmeltzerPRA2009}
B. Smeltzer, J. McIntyre, and L. Childress, Robust control of individual nuclear spins in diamond. {\it Phys. Rev. A} {\bf 80}, 050302 (2009).

\bibitem{NeumannScience2010}
P. Neumann \textit{et al.}, Single-Shot Readout of a Single Nuclear Spin. \textit{Science} \textbf{329}, 542 (2010).

\bibitem{Epstein2005} 
R. J. Epstein, F. M. Mendoza, Y. K. Kato, and D. D. Awschalom, Anisotropic interactions of a single spin and dark-spin spectroscopy in diamond. \textit{Nature Phys.} \textbf{1,} 94-98 (2005).

\bibitem{MaurerScience2012}
P. C. Maurer {\it et al.}, Room-temperature quantum bit memory exceeding one second. {\it Science} {\bf 336}, 1283-1286 (2012).

\bibitem{Kim:PRA1987}
M. S. Kim and P. L. Knight, Quantum-jump telegraph noise and macroscopic intensity fluctuations. {\it Phys. Rev. A} {\bf 36}, 5265 (1987). 

\bibitem{Molski:CPL2009}
A. Molski, Photon-counting distribution of fluorescence from a blinking molecule. {\it Chem. Phys. Lett.} {\bf 324}, 301-306 (2009).

\bibitem{Fuchs_PRL_2008}
G. D. Fuchs \textit{et al.}, Excited-State Spectroscopy Using Single Spin Manipulation in Diamond. \textit{Phys. Rev. Lett.} \textbf{101}, 117601 (2008).

\bibitem{GaliPRB2009}
A. Gali, Identification of individual $^{13}$C isotopes of nitrogen-vacancy center in diamond by combining the polarization studies of nuclear spins and first-principles calculations. {\it Phys. Rev. B} {\bf 80}, 241204 (2009).

\bibitem{Cohen}
C. Cohen-Tannoudji, B. Diu, and F.  Laloe, Quantum Mechanics, \textit{Wiley} (1977).

\bibitem{NR}
W. H. Press, S. A. Teukolsky, W. T. Vetterling, and B. P. Flannery,  Numerical Recipes - The Art of Scientific Computing. Ed. \textit{Cambridge University Press} (2007).


\end{thebibliography}
\end{document}